# The Design and Simulation of a Coarse-to-Fine Optical MEMS Accelerometer

**MOJTABA RAHIMI,**[1] **MAJID TAGHAVI,**[2] **AND MOHAMMAD MALEKMOHAMMAD**[1,*]

[1]*Department of Physics, University of Isfahan, Isfahan, Iran*
[2]*Laser and Plasma Research Institute, Shahid Beheshti University, Tehran, Iran*
*\*m.malekmohammad@sci.ui.ac.ir*

**Abstract:** In this paper, a novel coarse-to-fine optical MEMS accelerometer based on the Fabry–Pérot (FP) interferometer is proposed. The mechanical structure consists of a proof mass that is suspended by four L-shaped springs. The deflection of the proof mass due to the applied acceleration is detected using two FP cavities which comprise the optical system of the device. Using coarse-to-fine measurement and the dual wavelength method increases the sensitivity of the accelerometer as well as the linear measurement range simultaneously. The optical simulation shows that the sensitivity of the proposed device is 10 times as high as that of a similar optical MEMS accelerometer with one FP cavity. In addition, the proposed optical system is insensitive to the displacements of the proof mass in orthogonal directions as a result of which cross-axis sensitivity is considerably reduced. The minimum feature size of the structure is 15 μm and the optical signal is conducted completely through the optical fibers, facilitating the fabrication of the device. The simulation results are as follows: mechanical sensitivity of 190 nm/g, optical sensitivity of 8 nm/g, linear measurement of ±5 g, and first resonance frequency of 1141 Hz.

## 1. Introduction

Being a main sensing element in devices, MEMS (micro-electro-mechanical systems) accelerometers are widely employed in navigation systems, robotics, smart cellphones, and so on [1–5]. The typical working principle of these sensors is based on measuring the mechanical displacement of a proof mass due to the applied acceleration. Their small size, low power consumption, and available fabrication technology have put MEMS accelerometers at the center of attention. On the other hand, readout system complexity, non-immunity to electromagnetic waves, and sensitivity to thermal variations are the disadvantages of these sensors [6,7]. Incorporating MEMS and optics has led to the formation of the MOEMS technology which has a desirable performance in sensing acceleration due to its intrinsic immunity against electromagnetic interference, high resolution, and thermal stability. In MOEMS accelerometers, the mechanical displacement of the proof mass is measured by detecting the changes in the optical signal characteristics like intensity, frequency, and phase modulations [8–11]. Among these, frequency modulation has received the most attention because of its high accuracy and wide bandwidth. In this method, the applied acceleration causes a frequency shift in the optical spectrum which can be used as a gauge to determine the amplitude and direction of the acceleration. In the frequency modulation approach, acceleration can be measured based on the photonic band gap (PBG) effect [12,13] and by interferometric methods including Mach-Zehnder [14,15], Michelson [16], and Fabry–Pérot [17–20] interferometers.

    PBG MOEMS accelerometers consist of one-dimensional or two-dimensional photonic crystals [13,21] that are responsible for modulating the optical signal. A broadband optical signal passes through the PBG structure, leading to defect modes in the optical spectrum. Such defect modes are frequency sensitive to acceleration. Many attempts have been carried out to improve the performance of these accelerometers in terms of sensitivity [12,22], measurement range [21], and frequency [13]. However, they all have complex fabrication processes since

they use nanoscale structures like distributed Bragg reflectors, nanofingers, nanorings, and nanorods. Moreover, they have expensive and complex readout systems such as high-bandwidth light sources and high-resolution spectrum analyzers.

Among interferometric sensors, FP-based MOEMS accelerometers have a higher significance due to their advantages including high sensitivity to cavity length, very low cross-axis sensitivity, simple configuration, and straightforward fabrication in micron dimensions [23–25]. In FP-based accelerometers, the applied acceleration changes the length of the FP cavity. This causes an optical spectrum shift which can be used as a way to measure acceleration. As two important characteristics of FP-based accelerometers, measurement range and sensitivity are interdependent due to the repetitive pattern of the FP optical spectrum. Therefore, a wide measurement range leads to a low sensitivity and vice versa. In this paper, a novel FP-MOEMS accelerometer based on coarse-to-fine measurement is proposed. This accelerometer could overcome the limitations of previous accelerometers and provide a wide measurement range as well as a high sensitivity simultaneously. The proposed sensor has two FP cavities as coarse and fine measurement tools and has far better characteristics than those in previous studies.

The organization of this paper is as follows. In section 2, the working principles of the proposed accelerometer are discussed. In section 3, the optical and mechanical designs as well as the simulations are presented. Then, the results of the simulations are compared with those of some recent contributions in this field in section 4. Finally, the conclusions are given in section 5.

## 2. Working principles

The proposed accelerometer consists of a mechanical structure and an optical system. As shown in Fig. 1, the mechanical structure includes a proof mass that is attached to anchors by four L-shape springs. The structure is designed in a way that the proof mass can move in the Y-axis (the sensing axis) freely and its movements in the X and Z axes are negligible. As a result, the mechanical structure can be considered as a one-dimensional mass-spring system and its movements can be described as follows [26]:

$$m\ddot{y} + c\dot{y} + k_y y = -ma_y \tag{1}$$

where, m is the mass of the proof mass, $k_y$ is the equivalent spring constant of the structure, c is the damping coefficient, and y is the displacement from the rest position as a result of acceleration $a_y$. Similar to a mass-spring system, when an acceleration is applied to the structure in the Y-axis, the proof mass is deflected along the opposite direction. By measuring the value and direction of the proof mass displacement, the amplitude and direction of the applied acceleration can be determined.

One of the common methods for measuring the displacement of the proof mass in MOEMS accelerometers is using FP interferometers. In this method, the end of a cleaved single mode optical fiber (SMF-28) is placed near the proof mass (as shown in Fig. 1). As a result, an FP cavity is formed between the cleaved surface of the optical fiber and the proof mass cross-section. If an acceleration is applied to the sensor, the length of the FP cavity will be changed, leading to a shift in the FP cavity spectrum. As shown in Fig. 2-a, while acceleration in the +y direction leads to spectrum shifts to shorter wavelengths (blue shift), acceleration in –y direction leads to optical spectrum shifts to longer wavelengths (red shift). Therefore, by measuring the shift of the FP optical spectrum, the magnitude and direction of the applied acceleration can be determined.

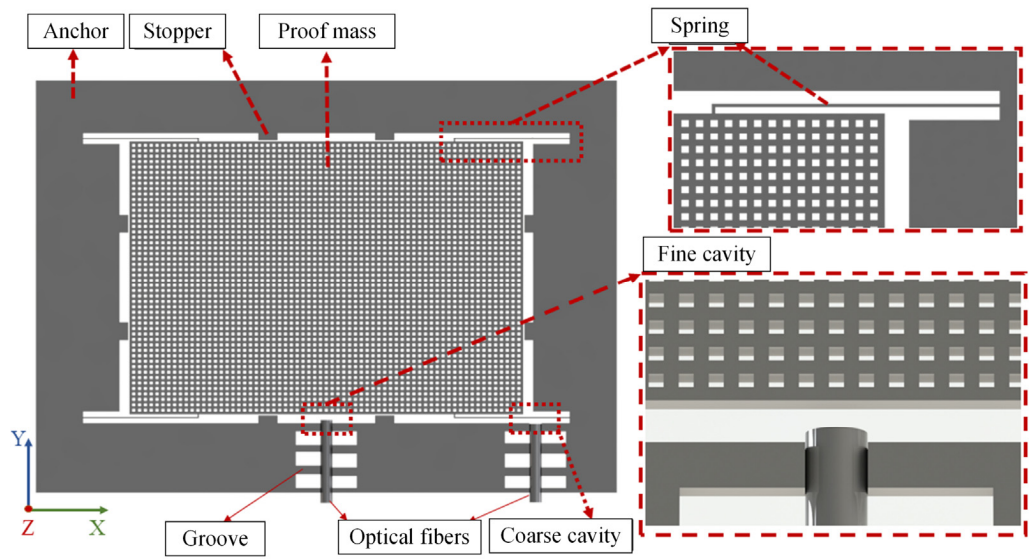

Fig. 1. The mechanical structure of the designed coarse-to-fine MOEMS accelerometer. The coarse and fine cavities, the proof mass, and the L-shaped springs are marked in the figure. Mechanical stoppers were used to prevent damage to the springs and proof mass by mechanical shocks and to restrict the movement of the proof mass in the X-axis. Square holes were cut in the proof mass to be used in the etching step of the fabrication procedure.

The optical spectrum shift can be measured using two different methods, namely wavelength monitoring and intensity monitoring. In the first one, the displacement of one spectrum peak can be measured using an optical spectrum analyzer. The main disadvantage of this method is that it requires an expensive and complex readout system. In practical applications, intensity monitoring, which is based on measuring the intensity variations of a specific wavelength using a simple power meter, is more preferable. Hence, this method was selected in the current paper. In this method, cavity length changes must lead to linear intensity variations. The range which provides a linear response in the FP optical spectrum is between two successive maximum and minimum spectra as shown in Fig. 2-b.

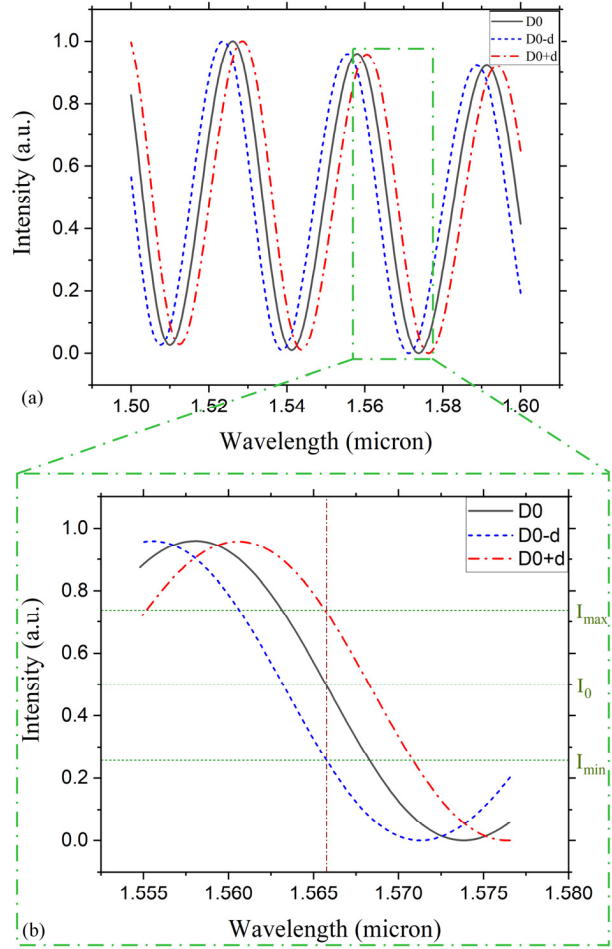

Fig. 2. (a) The Fabry-Pérot cavity spectrum and its blue and red shifts due to acceleration in the +y and -y directions, respectively; (b) the linear response range for a specific wavelength between two successive maximum and minimum spectra.

Optical sensitivity (the ratio of optical spectrum shifts to the applied acceleration ($\Delta\lambda/\Delta a$)) and measurement range ($\Delta a$) are two important characteristics of a MOEMS accelerometer. These two characteristics are interdependent in an FP-based accelerometer. Furthermore, in the intensity monitoring approach, the sensor needs to be designed so as to have a linear response in the measurement range. The value of the cavity length changes which provides a linear response for a specific wavelength is about 1/8 of the utilized wavelength. For example, the intensity changes at the wavelength of 1550 nm have a linear response when the cavity length changes are about 190 nm. Hence, for a specific measurement range ($\Delta a$), the maximum $\Delta\lambda$ is limited because of the linear response of the selected wavelength. Consequently, the wide acceleration measurement range reduces the sensitivity of the sensor and vice versa. This limitation imposes a tradeoff between the measurement range and the sensitivity of the sensor.

To overcome this significant limitation and achieve a high sensitivity as well as a wide acceleration measurement range, a novel coarse-to-fine MOEMS accelerometer (CFMA) based on two FP cavities is suggested. The first FP cavity (hereafter referred to as 'fine cavity') is formed between the single mode optical fiber and the proof mass cross-section. The second FP cavity (hereafter referred to as 'coarse cavity') is formed between another single mode optical fiber and the cross-section of one of the springs. Due to the stiffness of the spring, the coarse cavity has less length changes and less spectral shifts. Therefore, the coarse cavity provides a wide acceleration measurement range. Furthermore, changing the fine cavity length leads to a larger spectrum shift than that of the coarse cavity. Thus, a high sensitivity can be achieved by monitoring the optical spectrum shift in the fine cavity. A key point is that the mechanical structure should be designed in a way that the cavity length changes in the fine cavity become an integer multiple of the length changes in the coarse cavity. As a consequence, this relationship will also be true for their spectrum shifts. The fine-coarse ratio (α) is defined as follows:

$$\alpha = \Delta D_{fine} / \Delta D_{coarse} \tag{2}$$

where ΔD is the cavity length changes because of the applied acceleration. For example, if α=2, the maximum displacement and the spectral shift for the fine cavity will be twice as long as those of the coarse cavity for a specific acceleration. Therefore, this configuration will double the sensitivity of the sensor. As a result, the CFMA can measure acceleration with α-times higher sensitivity than a similar accelerometer with one cavity.

Another important issue to take into consideration is the linear response range in the fine cavity. The mechanical displacement in the fine cavity causes a spectrum shift which is larger than the linear response range. Hence, there are regions in which the sensor does not have a linear response to the applied acceleration. To deal with this challenge, a dual-wavelength approach for the readout system in the fine cavity is proposed. The criterion for choosing the two wavelengths is that the nonlinear response range of the first one matches the linear response range of the second one. This provides a continuous linear response range for the fine cavity.

## 3. Design and simulation

Designing the CFMA has two main steps. In the first step, the propagation of the light source in FP cavities is simulated and the optical spectra of the coarse and fine cavities are determined. In the next step, the mechanical structure is designed in a way that the applied acceleration to the structure creates the optical spectrum shift calculated in the first step.

### 3.1 Optical design

The optical system of the CFMA consists of coarse and fine FP cavities. The optical spectra of these two cavities are used as the readout system of the sensor. The end of the cleaved single mode optical fiber and the reflecting surface of the proof mass cross-section are the reflecting surfaces of the FP cavities as shown schematically in Fig. 3.

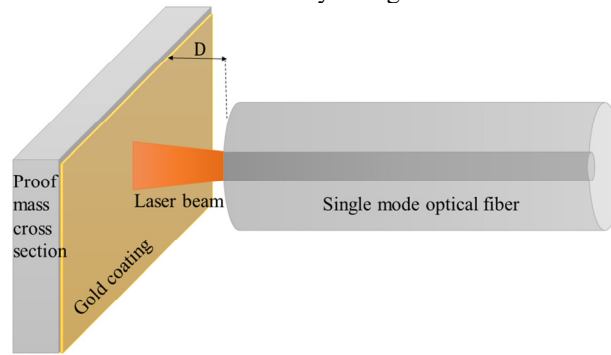

Fig. 3. The Fabry-Pérot cavity configuration in the CFMA. A gold coating was used to increase the reflectivity of the cavity surface.

The exiting light from the optical fiber experiences multiple reflections between the reflecting surfaces of the cavity which consequently leads to the creation of an optical interference spectrum. This configuration is used in the simulation of the Gaussian beam propagation in the coarse and fine cavities with the FDTD method. The parameters used in the simulation are listed in Table 1 and the output optical spectrum of the coarse cavity is shown in Fig. 4. The wavelength is selected as λ=1550 nm which is a commonly used wavelength in optical sensing. The coarse cavity parameters are selected in a way that the 1550 nm wavelength has a middle intensity in the rest position of the sensor. To determine the linear response range of this wavelength, intensity has been plotted versus cavity length in Fig. 5-a. This figure shows that the mechanical displacement of 190 nm (±95 nm for positive and negative accelerations) provides linear variations in the intensity of the selected wavelength (Fig. 5-b). This range determines the maximum mechanical displacement in the coarse cavity. Furthermore, the optical spectrum corresponding to the mechanical displacement of ±95 nm is plotted in Fig. 5-c. This figure also shows that the maximum spectral shift in the coarse cavity is 8 nm (±4 nm).

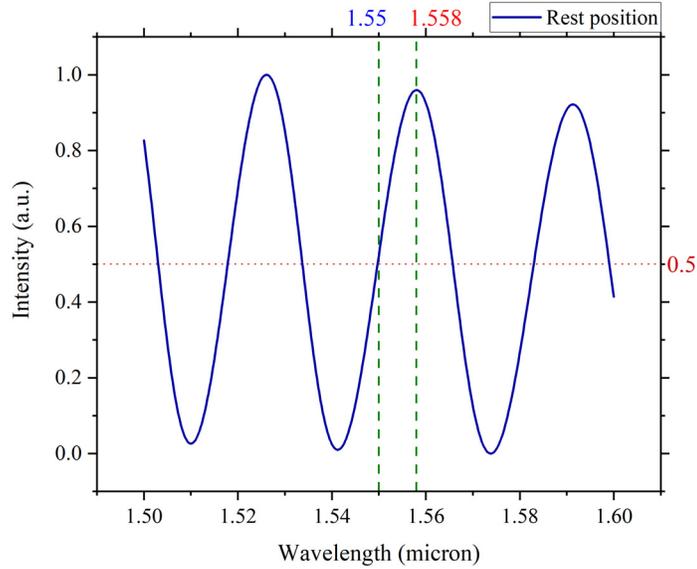

Fig. 4. The output optical spectrum for the simulated coarse cavity with the parameter mentioned in Table I. The first (1550 nm) and second (1558 nm) wavelengths are marked.

Table 1. The parameters used in the optical simulation

| Symbol | Description | Value |
|---|---|---|
| $D_{0\text{-coarse}}$ | Coarse cavity length | 36.30 μm |
| $D_{0\text{-fine}}$ | Fine cavity length | 36.58 μm |
| λ | Source wavelength | 1.5-1.6 μm |
| a | Optical fiber core diameter | 4.5 μm |
| $n_{co}$ | Optical fiber core refractive index | 1.446 |
| $n_{cl}$ | Optical fiber clade refractive index | 1.441 |
| R | Optical fiber surface reflection coefficient | 0.04 |
| T | Optical fiber surface transmission coefficient | 0.96 |
| n | Refractive index of the cavity medium | 1 |

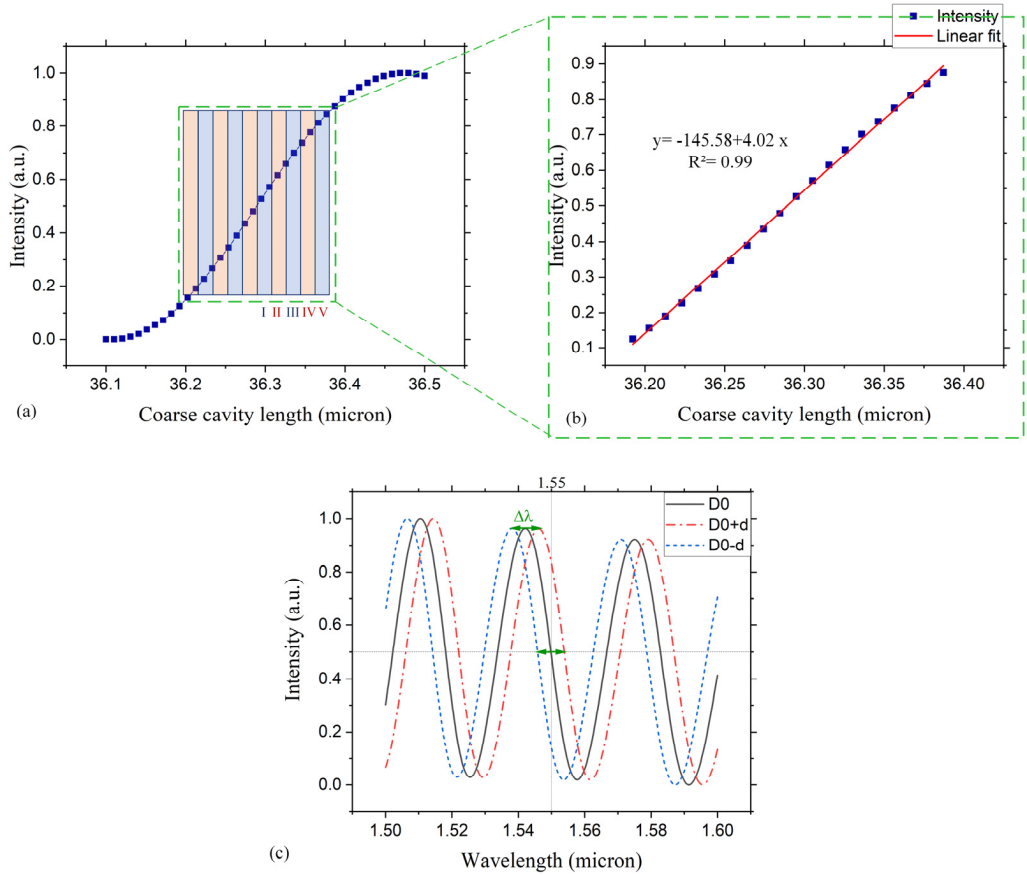

Fig. 5. (a) Intensity changes versus cavity length for the wavelength of 1550 nm in the coarse cavity; (b) the linear response range for the wavelength of 1550 nm; (c) the optical spectrum shifts corresponding to the mechanical displacements in the coarse cavity.

As mentioned in the previous section, the working principles of the CFMA are based on the wide acceleration measurement range provided by the coarse cavity and the high sensitivity provided by the fine cavity. The entire acceleration can be in the linear range of the coarse cavity because of the small changes in the coarse cavity length due to the applied acceleration. This paper aims to design a CFMA for measuring acceleration in ±5 g. Therefore, in ±5 g, the maximum mechanical displacement in the coarse cavity is ±95 nm.

The fine-coarse ratio for this accelerometer has been chosen as $\alpha=10$. Hence, the mechanical displacement of the fine cavity (±950 nm) is 10 times as large as that of the coarse cavity. This mechanical displacement can lead to a spectral shift out of the linear response range in the fine cavity spectrum, making the readout system inaccurate. To solve this problem, the dual wavelength approach is proposed for the readout system of the fine cavity. In this method, two optical signals with different wavelengths are sent to the fine cavity. The first wavelength is chosen similar to that of the coarse cavity signal at 1550 nm. The second wavelength is selected in a way that its linear response range completely matches the nonlinear response range of the first one. Considering the optical spectrum in Fig. 4, this condition is true for the wavelength at 1558 nm. To verify this condition, the output intensity has been plotted versus the fine cavity length for the first and second wavelengths in Fig. 6-a, while the linear response ranges of these wavelengths are shown in Fig. 6-b to f. It is obvious that a continuous linear response range in the fine cavity is achievable if the two optical signals are periodically monitored. When the coarse cavity length is in region I (shown Fig. 5-a), the intensity of the first optical signal should be monitored in the fine cavity (Fig. 6-b). For region II, the intensity of the second optical signal should be taken into account. This situation is correct for regions III to V and also for the opposite direction. To be more explicit, five regions are specified by different colors in Fig. 6-a. The amounts of the mechanical displacement and the spectrum shift as well as the related wavelengths for each region are presented in Table 2 by considering the amplitude and direction of the applied acceleration.

The coarse and fine cavities increase the maximum mechanical displacement to ±950 nm and the spectrum shift to ±40 nm. This is impossible in similar accelerometers with one simple cavity. Consequently, the CFMA has a mechanical sensitivity of 190 nm/g and an optical sensitivity of 8 nm/g in the acceleration measurement range of ±5 g. These values are 10 times as high as those of the MOEMS accelerometer with one FP cavity.

Table 2. Mechanical displacements, spectrum shifts, and readout wavelengths for different acceleration ranges (positive acceleration) in the fine and coarse cavities.

| Region | Acceleration range | Coarse cavity | | | Fine cavity | | |
|---|---|---|---|---|---|---|---|
| | | Mechanical displacement (nm) | Spectrum shift (nm) | Measurement wavelength | Mechanical displacement (nm) | Spectrum shift (nm) | Measurement wavelength |
| I | 0 to 1g | 0 to 19 | 0 to 0.8 | *First wavelength* | 0 to 190 | 0 to 8 | *First wavelength* |
| II | 1g to 2g | 19 to 38 | 0.8 to 1.6 | | 190 to 380 | 8 to 16 | *Second wavelength* |
| III | 2g to 3g | 38 to 57 | 1.6 to 2.4 | | 380 to 570 | 16 to 24 | *First wavelength* |
| IV | 3g to 4g | 57 to 76 | 2.4 to 3.2 | | 570 to 760 | 24 to 32 | *Second wavelength* |
| V | 4g to 5g | 76 to 95 | 3.2 to 4 | | 760 to 950 | 32 to 40 | *First wavelength* |

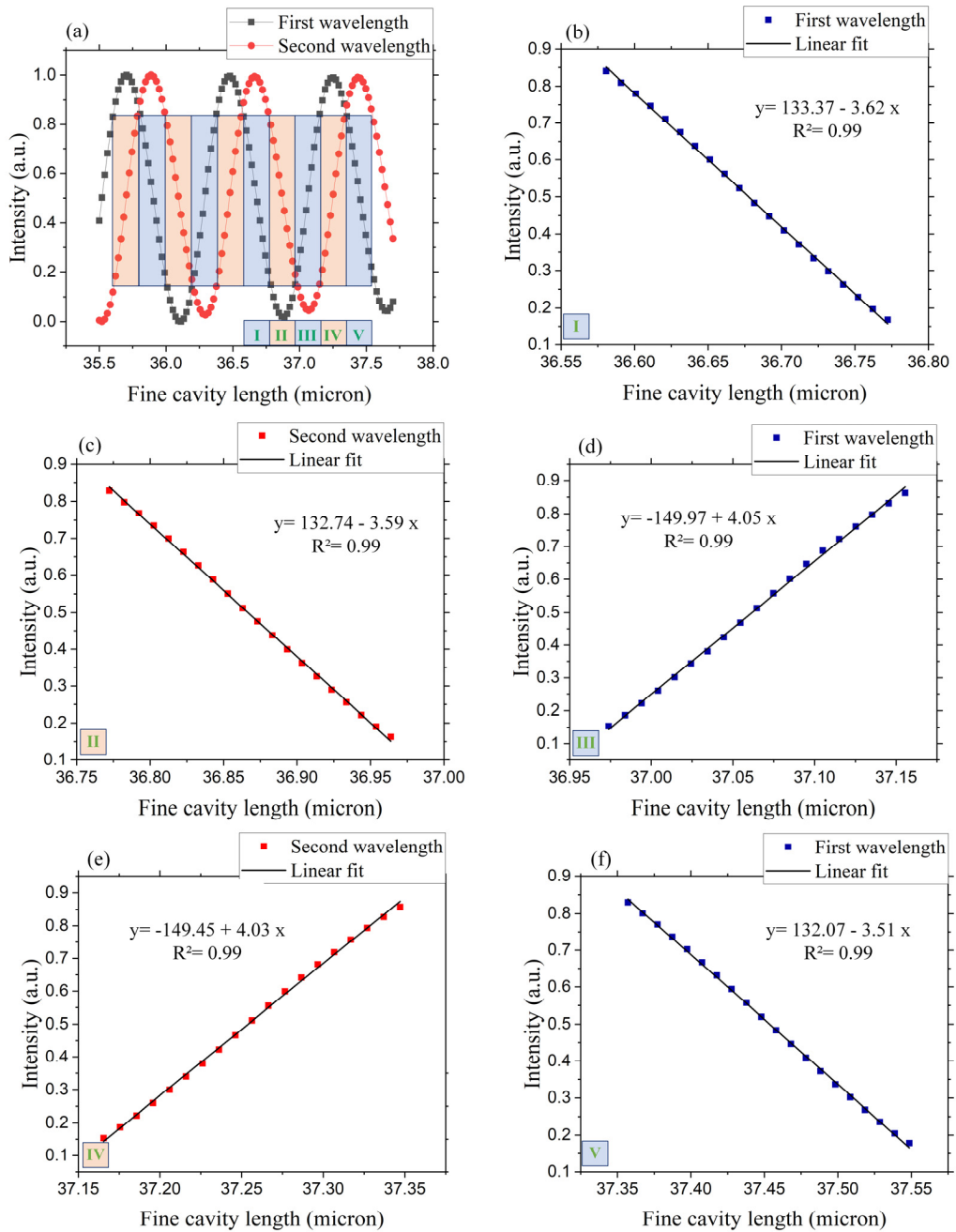

Fig. 6. (a) Intensity changes versus cavity length in the fine cavity for the first and second wavelengths; the linear regions for the first and second wavelengths are highlighted by two different colors; the linear regions of the fine cavity responses for the first and second wavelengths in regions I to V are respectively shown in diagrams (b) to (f).

## 3.2 Mechanical design

As shown schematically in Fig. 7, the structure of the accelerometer consists of a proof mass that is connected to anchors by four L-shaped springs. The stiffness of the springs must be calculated in a way that the applied acceleration leads to the movement of the proof mass. This movement must match the mechanical displacement calculated in the previous section. Equation (3) can be used to calculate the equivalent spring constant of the sensor structure in the sensing direction [26].

$$k_y = \frac{EtW_1^3(4L_1 + \beta L_2)}{L_1^3(L_1 + \beta L_2)}, \quad \beta = \left(W_1/W_2\right)^3 \tag{3}$$

where L1 and W1 are the length and width of the springs along the X-axis, respectively, L2 and W2 are the length and width of the springs along the Y-axis, respectively, t is the thickness of the springs, and E is the Young's modulus. The values of these parameters which are used in the mechanical simulation are mentioned in Table 3.

.

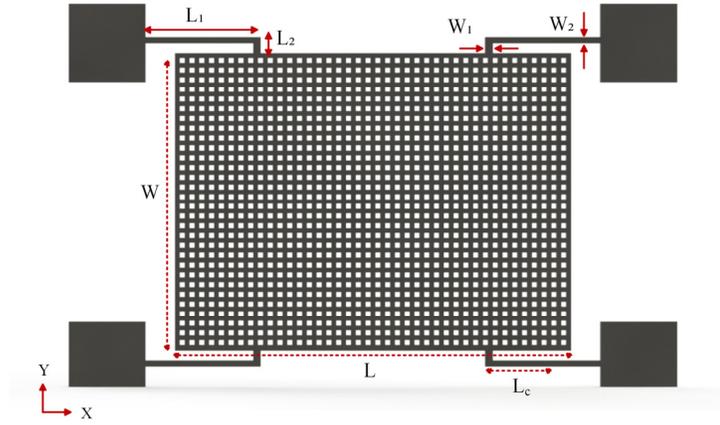

Fig. 7. The parameters used in the mechanical design are shown. The schematic structure of the sensor is not to the scale.

Table 3. The parameters used in the mechanical design and simulation

| Symbol | Description | Value |
|---|---|---|
| L | Proof mass length | 4020 μm |
| W | Proof mass width | 3020 μm |
| $L_1$ | Length of the springs along the X-axis | 1270 μm |
| $W_1$ | Width of the springs along the Y-axis | 15 μm |
| $L_2$ | Length of the springs along the Y-axis | 45 μm |
| $W_2$ | Width of the springs along the Y-axis | 15 μm |
| Lc | Coarse cavity location | 1010 μm |
| t | Thickness | 75 μm |
| E | Young's modulus of silicon | 170 GPa |
| ρ | Density of silicon | 2329 kg/m³ |

Resonance frequency is an important characteristic of the accelerometer that should be taken into account for determining the working bandwidth of the sensor. The first resonant frequency of the sensor could be approximated by the following equation [6]:

$$f_y = \frac{1}{2\pi}\sqrt{k_y/m} \qquad (4)$$

where m is the mass of the moving part of the sensor which is 1.59 mg for the designed structure. Using (3), the equivalent spring constant of the structure in the sensing direction is obtained as 81.87 N/m. Thus, the first resonant frequency for the proposed structure is about 1142 Hz.

The first optical fiber that forms the fine cavity is placed at the center of the proof mass (L/2). For the coarse cavity, the location of the optical fiber is $L_c$ where the displacement of the spring is 10 times lower than the displacement of the proof mass when an acceleration is applied. The mechanical simulations are performed in COMSOL Multiphysics and the results are summarized in Table 4. In addition, the mechanical displacements of the coarse and fine cavities due to the applied acceleration are shown in Fig. 8.

**Table 4. The results of the mechanical simulation (FEM).**

| Symbol | Description | FEM |
|---|---|---|
| $k_y$ | Spring constant along the Y-axis | 81.96 N/m |
| $k_x$ | Spring constant along the X-axis | $1.25\times10^6$ N/m |
| m | Mass of the proof mass | $1.59\times10^{-6}$ kg |
| $k_z$ | Spring constant along the Z-axis | $7.79\times10^4$ N/m |
| $f_y$ | Resonance frequency along the Y-axis | 1.141 kHz |
| $f_x$ | Resonance frequency along the X-axis | 141.5 kHz |
| $f_z$ | Resonance frequency along the Z-axis | 35.2 kHz |
| $S_y$ | Mechanical sensitivity along the Y-axis | 190.11 nm/g |
| $S_x$ | Mechanical sensitivity along the X-axis | 0.012 nm/g |
| $S_z$ | Mechanical sensitivity along the Z-axis | 0.2 nm/g |
| $C_{xy}$ | Cross-axis sensitivity between the X and Y-axes | 0.01 % |
| $C_{zy}$ | Cross-axis sensitivity between the Z and Y-axes | 0.1 % |

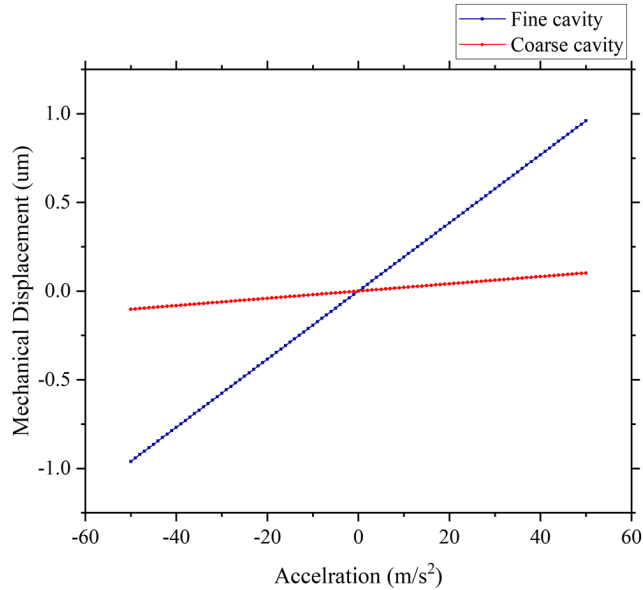

Fig. 8. The length changes in the coarse and fine cavities due to the applied acceleration.

L-shaped springs have a large spring constant in the X and Z directions (normal to the sensing axis). This reduces the cross-axis sensitivity of the sensor. The cross-axis sensitivities in the X ($C_{xy}$) and Z ($C_{zy}$) directions are calculated as follows:

$$C_{xy} = S_x/S_y, \qquad C_{xy} = S_z/S_y \qquad (5)$$

where $S_x$, $S_y$, and $S_z$ are the sensitivities along the X, Y, and Z directions, respectively. Hence, the cross-axis sensitivities between the Y and X and Y and Z directions are about 0.01% and 0.1%, respectively. However, the FP spectrum is not inherently sensitive to movement in perpendicular directions. Therefore, the cross-axis sensitivity of the sensor can be regarded as almost zero.

The frequency response of the sensor is simulated in COMSOL Multiphysics (FEM analysis) in order to specify the resonance modes of the sensor. The first four resonance modes of the structure are shown in Fig. 9. The first resonance mode is at 1141.1 Hz which is completely in the sensing direction (Y-axis) and is in good agreement with the value obtained by the theoretical equation. This frequency determines the working bandwidth of the sensor. The second mode is at 4419.1 Hz and oscillates in the perpendicular axis (Z-axis). The third and fourth modes are at 7729.8 and 7925.9 Hz, respectively, and cause the rotational movements of the proof mass. The frequencies of the second, third, and fourth modes are larger than that of the first resonance mode. Therefore, they do not interfere with the performance of the sensor.

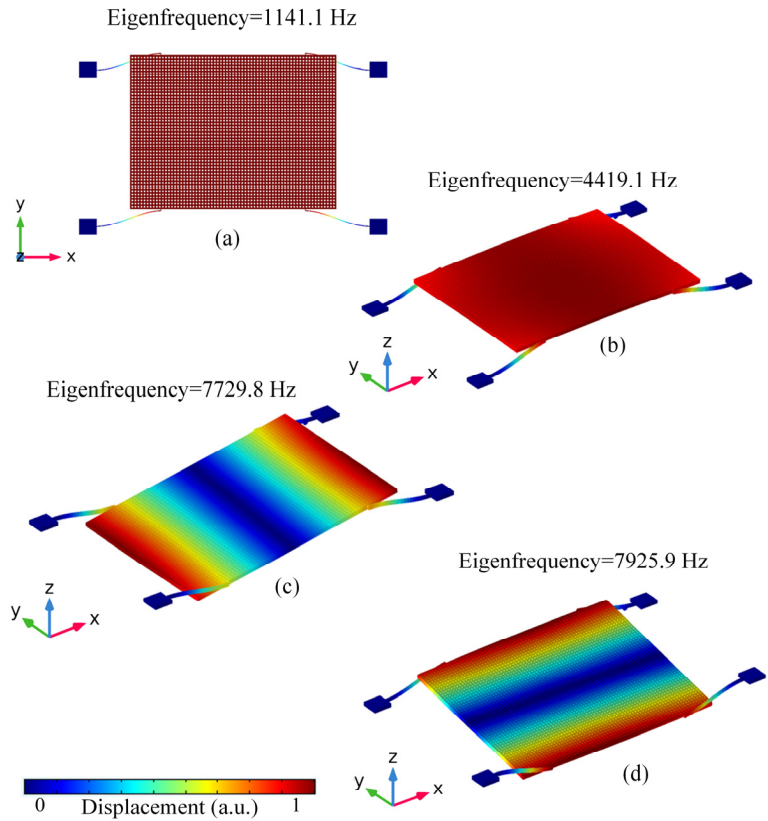

Fig. 9. The first four resonance modes of the CFMA simulated in COMSOL; (a) the first mode (1141.1 Hz); (b) the second mode (4419.1 Hz); (c) the third mode (7729.8 Hz); (d) the fourth mode (7925.9 Hz).

## 4. Analysis and comparison

In this section, the results of the design and simulation of the CFMA are compared with those of some important recent works in this field. For this purpose, the main characteristics of a MOEMS accelerometer are listed in Table 5.

One of the most important advantages of the designed accelerometer is the simplicity of the manufacturing technology compared to those of other works. The designed structure could be fabricated using simple and straightforward lithography processes [19]. In previous works, such structures as one- and two-dimensional photonic crystals were used which complicated the fabrication process. These difficulties become clear when the minimum feature sizes of the structures are considered. The minimum feature size in the designed CFMA is 15 μm which facilitates the fabrication process. Furthermore, in the designed CFMA, the optical signal is completely conducted through the optical fibers. Therefore, the problematic limitations of using waveguides are overcome.

Using an FP cavity makes the accelerometer insensitive to movements in the other two axes (X and Z) because no changes are created in the cavity spectrum due to displacements in these two axes. Moreover, the designed mechanical structure has a very low cross-axis sensitivity compared with other works. Consequently, the cross-axis sensitivity was reduced in the proposed device.

Considering the acceleration measurement range, the sensitivity of the proposed accelerometer is much higher than those reported in previous works. This is a significant characteristic of the proposed sensor. Thus, the overall performance of the proposed sensor is desirable.

Table 5. Comparing the main characteristics of the designed CFMA with those of the MOEMS accelerometers reported in some recent works.

| Characteristics | CFMA | 2020 [22] | 2020 [13] | 2019 [27] | 2019 [23] | 2017 [21] |
|---|---|---|---|---|---|---|
| Measurement range | ±5 g | ±0.517 g | ±217 g | ±1.56 g | ±1 g | ±156 g |
| Mechanical sensitivity ($\Delta x/\Delta a$) | 190 nm/g | 242 nm/g | 0.612 nm/g | 781.64 nm/g | 70 nm/g | 1.6 nm/g |
| Optical sensitivity ($\Delta\lambda/\Delta a$) | 8 nm/g | 1017 nm/g | 2.07 nm/g | 964.35 nm/g | 3.2 nm/g | 0.0756 nm/g |
| Resonance frequency | 1141 Hz | 1037 Hz | 20911 Hz | 562.85 Hz | 1676 Hz | 12935 Hz |
| Footprint of the proof mass | 4020 μm × 3020 μm | 2000 μm × 2000 μm | 116 μm × 116 μm | 4207 μm × 4207 μm | 8000 μm × 6000 μm | 200 μm × 200 μm |
| Cross-axis sensitivity | 0.01% (X) 0.1% (Z) | 9.67% (X) 2.6% (Z) | 0.86% (X) 0.96% (Z) | 7% (X) 1% (Z) | 1.02% (X) 1.36% (Z) | Not mentioned |
| Main structure | Fabry-Pérot cavity | 1D photonic crystal | 1D photonic crystal | 1D photonic crystal | Fabry-Pérot cavity | 2D photonic crystal |
| Minimum feature size | 15 μm | 109.8 nm | 109.8 nm | 109.8 nm | 70 μm | 240 nm |
| Fabrication complexity | Simple | Difficult | Difficult | Difficult | Simple | Difficult |

## 5. Conclusion

In this paper, an optical MEMS accelerometer was designed based on the FP interferometer. The readout system of the proposed accelerometer was based on coarse-to-fine measurement methods. The proposed approach provided a wide measurement range and a high sensitivity. The optical system consisted of two FP cavities and the mechanical system was composed of a proof mass and springs. These systems were simulated using the FDTD and FEA methods. There was no nanoscale feature in the designed structure. Hence, the sensor was entirely compatible with micro-fabrication technology. The optical sensing of the proposed device was based on measuring the output power of the two optical signals and needed neither an expensive broadband light source nor a high-resolution spectrum analyzer. According to the simulations, the designed accelerometer provided the optical sensitivity of 8 nm/g and the mechanical sensitivity of 190 nm/g in the linear measurement range of ±5 g. Furthermore, the first resonance frequency of the designed accelerometer was 1141 Hz. These characteristics make the proposed device desirable for many applications including inertial navigation and aerospace. The fabrication and characterization of the proposed device are still ongoing and will be the subject of our next paper.

**Disclosures.** The authors declare no conflicts of interest.

**Data availability.** Data underlying the results presented in this paper are not publicly available at this time but may be obtained from the authors upon reasonable request.